\documentclass[letterpaper,14pt]{article}
\usepackage{setspace}
\doublespace
\usepackage{cite}
\usepackage{amsmath}
\usepackage{amsfonts}
\usepackage{amssymb,multirow,wrapfig}
\usepackage{color}
\usepackage{xfrac}
\usepackage{xcolor}
\usepackage{graphicx}
\colorlet{shadecolor}{blue!20}
\usepackage{framed}

\usepackage{epstopdf}

\usepackage[small,normal,up]{caption2}

\newcommand{\be}[0]{\begin{equation}}
\newcommand{\ee}[0]{\end{equation}}
\newcommand{\bea}[0]{\begin{eqnarray}}
\newcommand{\eea}[0]{\end{eqnarray}}

\begin{document}

\setcounter{secnumdepth}{0}

\begin{center}

\LARGE\textbf{Efficient separation of the orbital angular momentum eigenstates of light}

\vspace{4mm} \normalsize\textbf{Mohammad Mirhosseini$^{1,*}$, Mehul Malik$^{1,2}$, Zhimin Shi$^{1,3}$, and Robert W. Boyd$^{1,4}$}

\vspace{1mm}
\footnotesize
\textit{$^1$The Institute of Optics, University of Rochester,
Rochester, New York 14627 USA}\\
\textit{$^2$Institute for Quantum Optics and Quantum Information (IQOQI), Austrian Academy of Sciences, Boltzmanngasse 3, A-1090 Vienna, Austria.}\\
\textit{$^3$Department of Physics, University of South Florida, Tampa, FL 33620, USA}\\
\textit{$^4$Department of Physics, University of Ottawa, Ottawa, ON K1N 6N5 Canada}\\
\textit{$^*$mirhosse@optics.rochester.edu}

\end{center}
\vspace{4mm}

\textbf{Orbital angular momentum (OAM) of light is an attractive degree of freedom for fundamentals studies in quantum mechanics. In addition, the discrete unbounded state-space of OAM has been used to enhance classical and quantum communications. Unambiguous measurement of OAM is a key part of all such experiments. However, state-of-the-art methods for separating single photons carrying a large number of different OAM values are limited to a theoretical separation efficiency of about 77 percent. Here we demonstrate a method which uses a series of unitary optical transformations to enable the measurement of lightÕs OAM with an experimental separation efficiency of more than 92 percent. Further, we demonstrate the separation of modes in the angular position basis, which is mutually unbiased with respect to the OAM basis. The high degree of certainty achieved by our method makes it particularly attractive for enhancing the information capacity of multi-level quantum cryptography systems.}

 \vspace{4mm} \noindent\normalsize\textbf{Introduction}
\vspace{4mm}

It has long been known that the polarization state of the optical field can be associated with the spin angular momentum \cite{Poynting, Beth}. In 1992, it was shown by Allen et al. that in the paraxial regime, a  \(e^{i\ell\varphi}\) vortex phase structure of a circularly symmetric beam corresponds to \(\ell \hbar\) units of OAM \cite{Allen}. Since then, OAM of light has been found as a useful tool in a variety of applications \cite{Zeilinger, Molina, Malik2, Barreiro, Willner}. OAM modes with different \(\ell\) values form a large orthonormal set of functions that can be used to encode information. It has been suggested that OAM encoding can be used alongside polarization to increase the channel capacity of communication systems \cite{BoydSPIE, Malik}. Additionally, the use of a multi-level encoding basis such as the OAM basis can increase the tolerance of QKD systems to eavesdropping \cite{Bourennane}. Clearly, applications of this sort will require a method of sorting photons carrying OAM with a high separation efficiency, which is a measure for the degree of separation of the sorted OAM modes, as well as a high power transmission efficiency, which is an indicator of the optical throughput of the system in presence of loss. Although the two eigenstates of polarization can be separated easily by using a polarizing beam splitter, separation of OAM eigenstates has proven to be challenging until recently \cite{Leach, Berkhout, Lavery}. 

A forked hologram can be used to add or remove spiral phase structures in its diffraction orders \cite{Gibson}.\,Similarly, a q-plate can be used to create OAM modes by utilizing the spin-orbit coupling in inhomogeneous anisotropic media \cite{Marrucci, Nagali, Marrucci2}. Such elements can be used for measuring the OAM content of a beam by performing a series of projection measurements \cite{Zeilinger, Willner}. In this method, an input laser beam illuminates multiple forked holograms or q-plates with different quantum numbers and the transmitted light is collected by multiple single mode fibers which correspond to different OAM values. If an OAM beam with quantum number  \(\ell\) enters this system, its phase structure will only be corrected by the element that adds a phase of \(- \ell\) and a detection happens at the corresponding detector. This method can be used to distinguish between different OAM modes with high accuracy \cite{Karimi, Willner}. However, projection measurements are always limited by a success rate of \(1/N\), where \(N\) is the number of the OAM modes involved in the experiment.  This property makes this method unsuitable for single photon experiments that involve a large number of OAM modes. 

More sophisticated holograms designs can be made to achieve separation of several OAM modes using a single phase-screen \cite{Gibson}. However, none of these designs have proven to be efficient enough for separation of OAM modes at the single-photon level.  Similarly, the results achieved using thick holograms have been only marginally better \cite{Gruneisen}. Leach et al. have presented a technique for separating OAM modes at the single-photon level based on using a Mach-Zehdner interferometer with a Dove prism in each arm \cite{Leach}. This setup allows the separation of single photons based on their parity. Although in principle this method works with a 100\% efficiency, separating \(N\) modes requires \(N-1\) cascaded interferometers. The challenge in designing and aligning such a complicated system considerably limits its applications. 

Berkhout et al. have recently demonstrated a very successful method for sorting OAM modes that employs a Cartesian to log-polar transformation \cite{Berkhout}. Under this transformation, the azimuthal phase profile of an OAM mode is mapped to a tilted planar wavefront. As a result, a set of OAM modes with different quantum numbers \(\ell\) is converted to a set of truncated plane waves with wavefront tilts which are proportional to \(\ell\). These plane waves are then separated by a single lens. This method of sorting OAM modes is exclusively based on unitary transformations and in theory can achieve a power transmission efficiency of unity. However, the degree of separation of two neighboring OAM modes is limited by diffraction due to the finite size of the unwrapped modes. 

In this work we experimentally implement an architecture that combines log-polar coordinate mapping with refractive beam copying to substantially enhance the separation of the transformed OAM modes. Furthermore, we use a variation of this method for sorting angular(ANG) modes, which form an unbiased basis set with respect to the OAM modes. We achieve a separation efficiency of \(92.1 \pm 0.7\) for a set of 25 OAM modes and \(92.7 \pm 0.5\) for the corresponding 25 ANG modes, which are equivalent to mutual information values of 4.18 and 4.16 bits per detected photons respectively.

\vspace{4mm} \noindent\normalsize\textbf{Results}
\vspace{4mm}

\textbf{Mode overlap in log-polar coordinate mapping.} To achieve the coordinate transformation, the beam's amplitude \(a(x,y)\) is multiplied by a phase-only transmittance function \(\phi_1(x,y)\). The amplitude of the field in the conjugate Fourier plane can be calculated by using the stationary phase approximation, resulting in a geometrical map between the input and output planes \cite{Bryng-JOSA}. This mapping is accompanied by an undesired varying phase term. This phase distortion arises from variations in the optical path length and can be removed by introducing a second phase plate \(\phi_2(u,v)\) at the Fourier plane \((u,v)\) of the input plane. A coordinate mapping of \(v = a \arctan(y/x)\) and \(u = -a\ln (\sqrt{x^2+y^2}/b)\) can be achieved using the following phase elements \cite{Saito,Berkhout}

\begin{align}\label{eq:Elements}
\phi_1(x,y)& = \frac{2\pi a}{\lambda f}\left[y\arctan(\frac{y}{x}) - x\ln(\frac{\sqrt{x^2+y^2}}{b}) + x\right], \\
\phi_2(u,v)& = -\frac{2\pi a}{\lambda f}\exp(-\frac{u}{a})\cos(\frac{v}{a}). 
\end{align}

\begin{figure}[t!]
\centerline{\includegraphics[scale= 0.75]{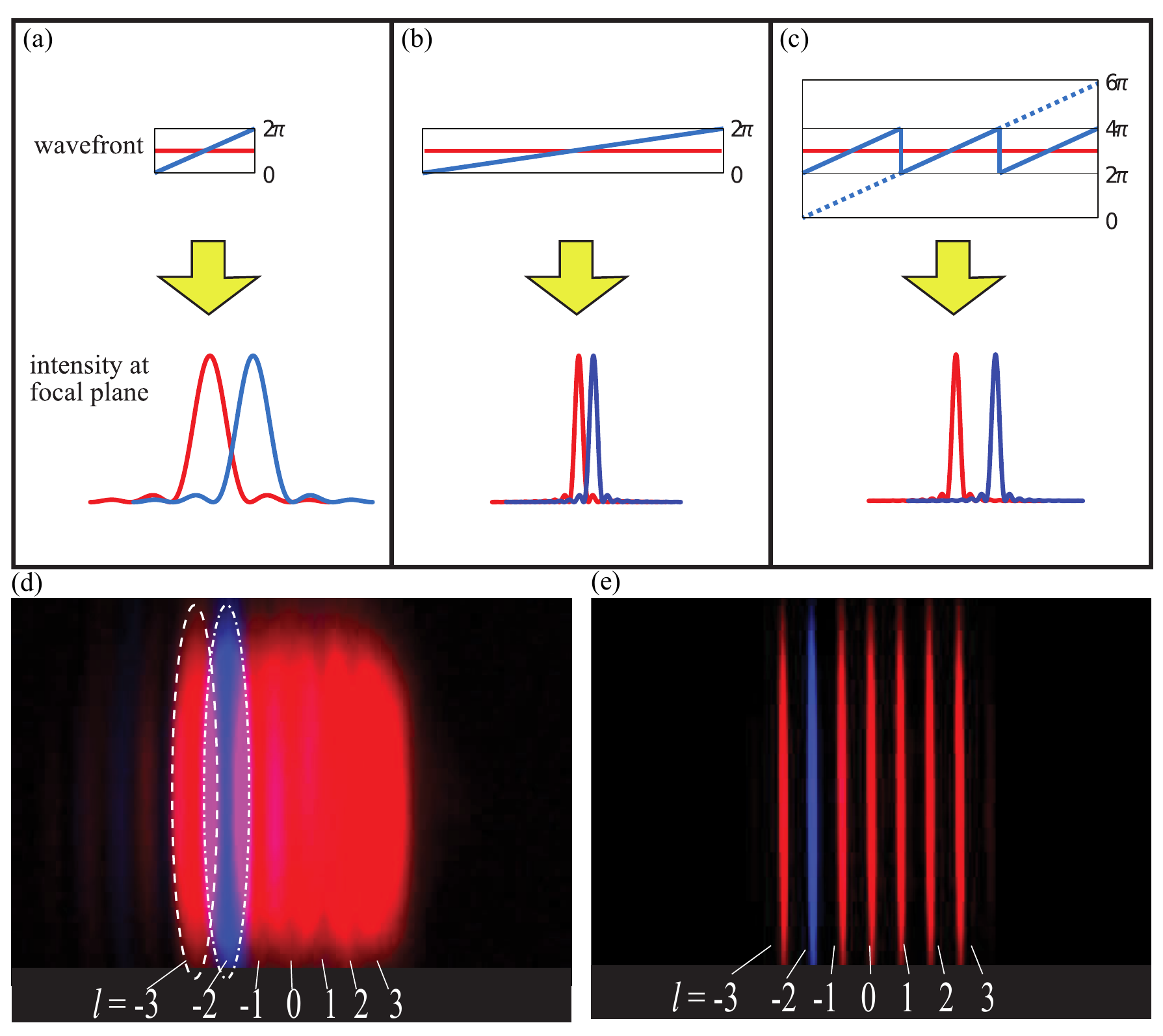}}
\caption{\textbf{Effect of beam copying on mode overlap.} a) The tilted plane waves, resulting from the transformation of an OAM modes are focused to overlapping spots by a lens. b) Magnified plane waves can be focused to narrower spots, which are spaced closer to each other. c) Multiple coherent copies of the transformed plane waves are placed next to each other. d) Experimental data for OAM modes focused after the log-polar mapping. The data for seven modes  \(\ell = -3:3\) are super-imposed on top of each other. The \(\ell = -2\) mode is displayed in false color to mark the overlap with the neighboring modes. e) Experimental data for the modes from part (d) when enhanced by making seven copies in the Fourier domain.}
\label{fig:Sinc}
\end{figure}

It is easy to check that the azimuthal phase profile of an OAM mode is mapped to a tilted planar wavefront under this transformation. The set of plane waves resulting from the transformation of OAM modes with different quantum numbers \(\ell\) are then separated by a single lens positioned in the \((u,v)\) plane. 

Fig.\,\ref{fig:Sinc}(a) demonstrate the situation where the plane waves resulting from the transformation of two neighboring OAM modes are focused to diffraction-limited spots which are partially overlapping. In Ref.\ \cite{Lavery}, the authors have measured the fraction of intensity overlap between the neighboring modes to be about 20 percent (See Fig.\,\ref{fig:Sinc}(e)). This is in fact an inherent limitation of this method and cannot be avoided by magnifying the transformed plane waves. Magnifying a plane wave results in a decrease in the tilt angle across its wavefront. Such magnified plane waves can be focused to narrower spots, but the spacing between the spots gets reduced accordingly such that the degree of overlap between them remains constant (Fig.\,\ref{fig:Sinc}(b)).

Motivated by a suggestion of Berkhout et al \cite{Berkhout}, we recently investigated the effects of a beam-copying device on the degree of separation of transformed OAM modes \cite{Collin}. Our previous simulation work along with the preliminary experimental results suggested that the use of such a device may dramatically reduce mode overlap. In the present paper, we describe the full implementation of this idea. This method works by mapping each OAM mode to multiple copies of the tilted planar wavefronts. In this situation the width of the transformed beam has been increased, while the angle of the tilt across it remains constant. Note that the periodic phase jumps between the neighboring copies of the truncated plane waves are equal to integer multiples of \(2\pi\), so the resulting larger beam will have a smooth wavefront. These converted modes can now be focused to a series of spots which have the same spacing as before but are much narrower in width \cite{Collin} (Fig.\,\ref{fig:Sinc}(c) and Fig.\,\ref{fig:Sinc}(e)).

\begin{figure}[t!]
\centerline{\includegraphics[scale= 0.75]{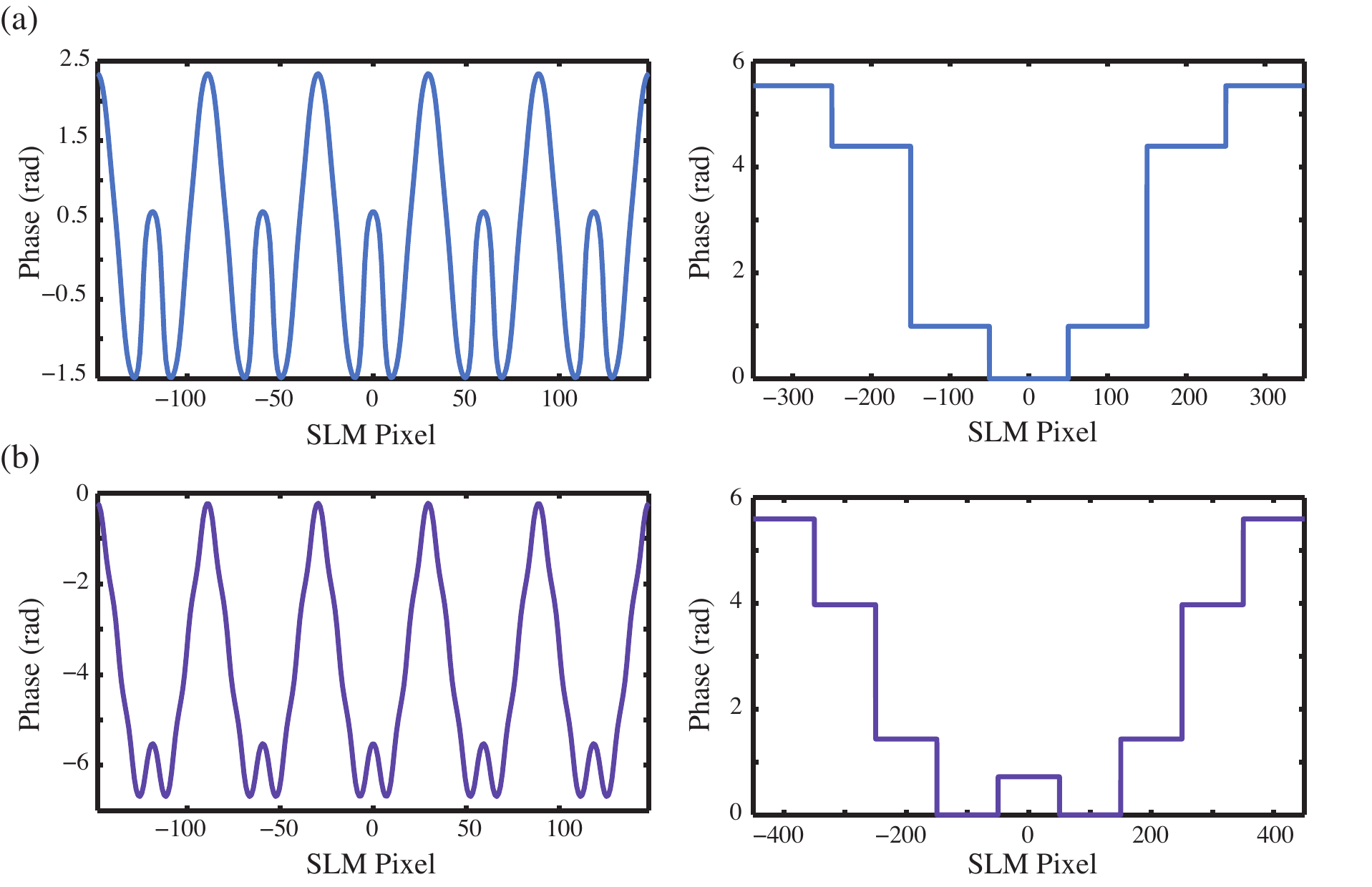}}
\caption{\textbf{Holograms for diffractive beam copying.} (a) The hologram for creating 7 copies (shown on the left) along with the corresponding hologram for correcting the relative phase between the copies (shown on the right). (b) The fan-out and phase corrector holograms for creating 9 copies. }
\label{fig:FanOutPhase}
\end{figure} 

\textbf{Refractive beam copying.} We have used a periodic phase-only hologram known as a fan-out element to make multiple coherent copies of the unwrapped OAM modes. The phase structure of such a device can be described as \cite{Romero} 
\begin{equation}
\Psi_{2N+1}(x) = \tan^{-1} \left(\frac{\sum\limits_{m=-N}^{N}\gamma_m \sin[(2\pi s/\lambda)mx+\alpha_m]}{\sum\limits_{m=-N}^{N}\gamma_m \cos[(2\pi s/\lambda)mx+\alpha_m]}\right).
\end{equation}
where \(2N+1\) is the number of copies of the beam, \(s\) is the angular separation between them, and \(x\) is the transverse dimension along which the copies are made (See Fig. \ref{fig:FanOutPhase} (a) and (b)). Here, \(\gamma_m\) and \(\alpha_m\) are relative phase and intensity parameters associated with different diffraction orders (See Supplementary Note 1 and Supplementary Table S1). These parameters can be optimized using numerical or analytical methods to uniformly distribute more than 99 percent of the incident light between the copies \cite{Romero, Prongue}. 

 \begin{figure}[b!]
\centerline{\includegraphics[scale= 0.5]{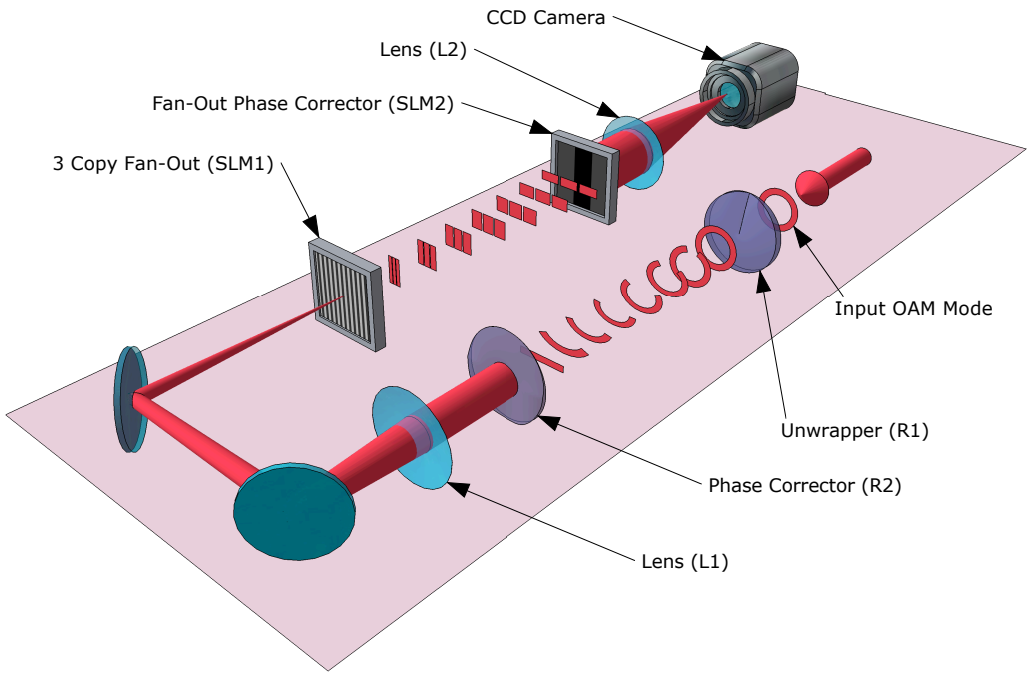}}
\caption{ \textbf{The schematic of the mode sorting setup.} The optical transformation converts the OAM modes to plane waves as they go through the first and second refractive phase elements.  An SLM is used to create multiple copies of the unwrapped beam. A lens focuses the resulting wide beam into a spot, after the phase distortions are removed by the second SLM. (To simplify the demonstration, a fan-out element creating only 3 copies has been displayed. We have also eliminated the Fourier transform lens between the fan-out element and the phase corrector, for further simplification.)  }
\label{fig:Setup}
\end{figure} 

\textbf{Sorting OAM modes.} In our experiment, we generate OAM modes using a spatial light modulator (SLM) and a 4f system of lenses \cite{Arrizon}. As shown in Fig.\,\ref{fig:Setup} the OAM beam then propagates through the refractive elements which perform the log-polar coordinate transformation \cite{Lavery}. A fan-out element realized on SLM1 generates multiple copies of this truncated plane wave mode.  For clarity Fig.\,\ref{fig:Setup} shows a fan-out element only creating 3 copies. However, we have used a fan-out design which creates 7 copies of the beam \cite{Romero} (See Supplementary Fig. S1 and Supplementary Table S1). SLM2 is used to correct the relative phase between these copies at the Fourier plane of the fan-out element. The holograms for the fan-out element and the corresponding phase corrector are shown in Fig.\,\ref{fig:FanOutPhase} (a). Finally, a lens is used to focus the extended plane-wave to a spot, where a charge-coupled-device (CCD) camera measures its intensity profile.

\begin{figure}[t!]
\centerline{\includegraphics[scale= 0.80]{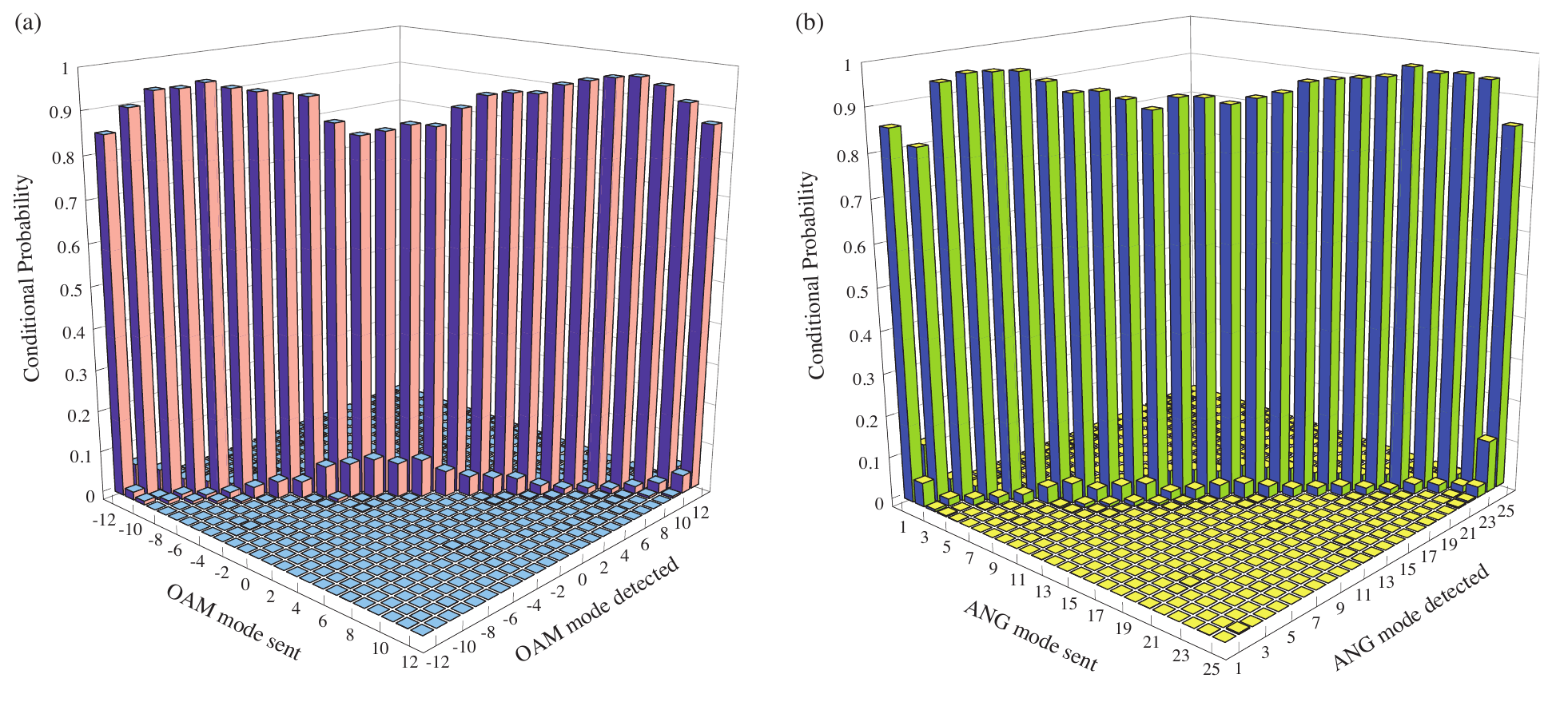}}
\caption{\textbf{The conditional probability of detection.} The horizontal axes indicate the indices for the sent and received (a) OAM modes and (b) ANG modes. For spatial profile of the modes see Supplementary Fig. S3 and Supplementary Fig. S4. The details of calculations are presented in the Supplementary Note 3.}
\label{fig:Results}
\end{figure}

 We have measured OAM modes with quantum numbers of \(\ell =\)-12 to \(\ell =\)12 using our method. We divide the area on the CCD into 25 non-overlapping adjacent spatial bins. Each of the bins corresponds to the central position of the spot resulting from the transformation of an OAM mode. By sending a known OAM mode through the system and measuring the power in all 25 bins, we have constructed the conditional probability matrix for detection of all 25 OAM modes (Fig.\,\ref{fig:Results}(a)). The average probability of correctly detecting an OAM mode is calculated to be \(92.1\pm 0.7\) percent. This is slightly lower than the theoretical maximum of 97 percent \cite{Collin} due to the non-ideal behavior of a phase-only fan-out element, undesired effects from pixelation of the SLMs, and small misalignments in the system. However, this is substantially higher than the maximum separation efficiency of 77 percent achieved by just the log-polar mapping method \cite{Berkhout}. We have compared these two methods quantitatively in the Supplementary Note 2.

Another common measure of quantizing the separation efficiency of a mode sorter is the mutual information between the set of sent and detected modes. For the data presented here, the mutual information is calculated to be 4.18 bits per detected photon, as compared to the theoretical upper limit of 4.64 bits per detected photon for a set of 25 modes (See Supplementary Note 3 and Supplementary Fig. S2). It should be emphasized that these values are calculated for the detected photons and they exclude the technical loss in the experimental setup. The use of this metric can be justified considering the fact that the losses are caused by the limitations in technology and they are not inherent to our method. The refractive elements that perform the log-polar mapping have a combined transmission efficiency of 85 percent \cite{Lavery}. The SLMs that realize the fan-out element and its phase corrector are each measured to have a diffraction efficiency of 42 percent (For a complete analysis of the effect of loss on the overall information capacity of the system, please see Supplementary Note 3).

\begin{figure}[t!]
\centerline{\includegraphics[scale= 0.65]{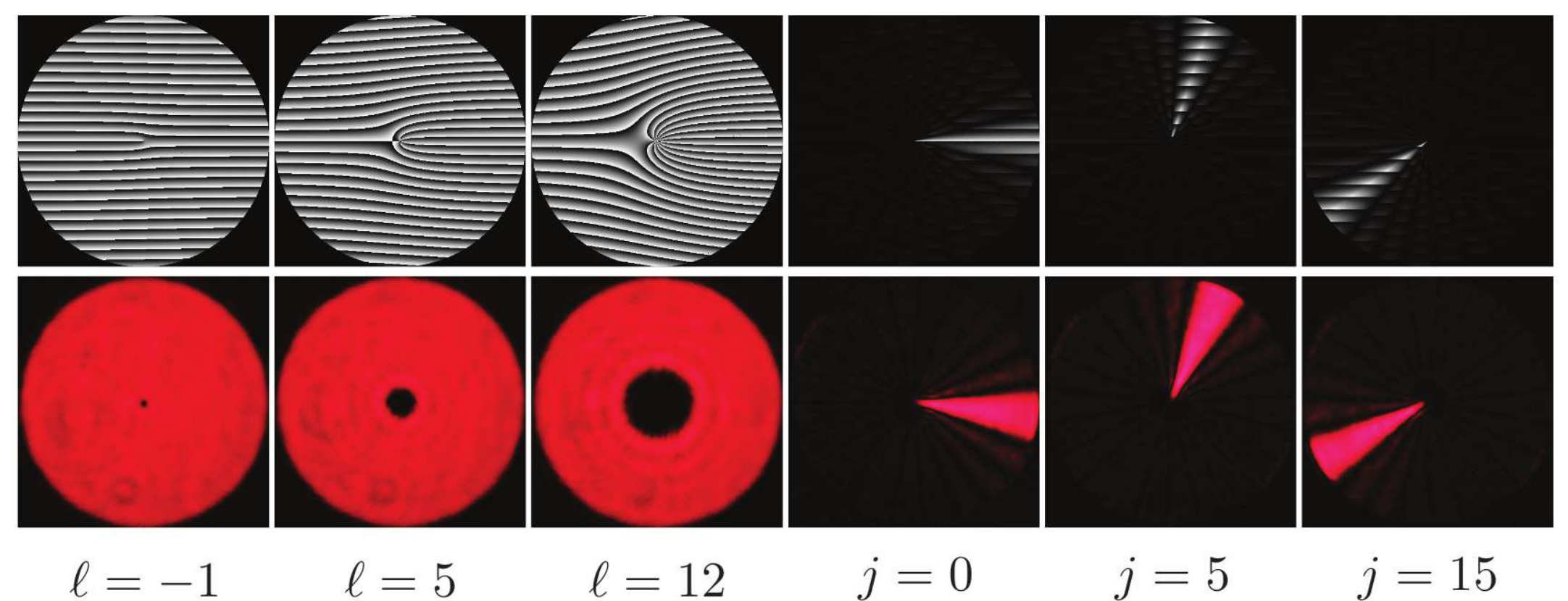}}
\caption{\textbf{Generation of OAM and ANG modes.} Examples of holograms (top line) for generating OAM and ANG modes along with the intensity profile (bottom line) of the resulting spatial modes}
\label{fig:SampleModes}
\end{figure}

\textbf{QKD and angular modes.} In order to guarantee security against eavesdropping in QKD, at least two mutually unbiased bases (MUBs) are required \cite{Bennett}. A MUB for OAM can be constructed from the superposition of OAM modes with equal weights and a relative phase \cite{Malik, Collin}:
\begin{equation}
\label{eq:angMode}
|\theta_j\rangle= \frac{1}{\sqrt{2L+1}} \sum_{\ell=-L}^{L} |\ell\rangle \: e^{-i 2 \pi j\ell/(2L+1)}.
\end{equation}
Here, an OAM mode is written as \( |\ell\rangle\) and \(L\) is the maximum OAM quantum number which is used in forming the basis. Members of this MUB are referred to as the angular (ANG) modes, due to angular localization of their intensity patterns (Fig.\,\ref{fig:SampleModes}). It is easy to check that if ANG modes are measured in the OAM basis, all the OAM modes in the range \( |\ell| \leq L\) may be detected with equal probabilities. ANG modes with different indices are orthogonal in term of their amplitudes, however, the intensity profiles of these modes have side lobes around their central angular component which creates substantial overlap between any two neighboring modes (See Fig.\,\ref{fig:SampleModes}). A detection scheme based on the intensity confinement of ANG mode has the rate of error of 23 percent \cite{Collin}.

\textbf{Sorting ANG modes.} An efficient method of sorting both OAM and ANG modes is needed to realize an OAM-based QKD system. Previous methods of sorting OAM modes are incapable of sorting ANG modes \cite{Malik}. However, a variation of the sorting method described above can be used to sort these modes \cite{Collin}. Since the unwrapped OAM and ANG modes are related via a Fourier transform in the horizontal direction, we can sort ANG modes by removing the Fourier-transforming lens L1 after the second refractive element R2. Instead the unwrapped mode is imaged directly into the fan-out element.  We use a fan-out element creating 9 copies (See Supplementary Note 1 and Supplementary Table S1). The holograms for the fan-out element and the corresponding phase corrector are shown in Fig.\,\ref{fig:FanOutPhase}(b). The ANG modes are transformed to a series of separated spots at the focal plane of the lens L2.

The conditional probability matrix for 25 ANG modes is shown in Fig.\,\ref{fig:Results}(b). The average probability for correct detection of an ANG mode in this set is measured to be \(92.7\pm 0.6\) percent. The mutual information calculated from the measured conditional probability matrix equals 4.16 bits per detected photon, demonstrating the capability of sorting ANG modes with approximately the same separation efficiency as that of the OAM modes. 

In this experiment, we have demonstrated a method for unambiguous discrimination of OAM modes. A complex optical transformation is used along with a refractive beam-copying method to map OAM modes to a set of extended plane waves which are then separated using a lens. Using our technique, we have measured a channel capacity of 4.18 bits per detected photon for a free-space communication system using 25 OAM modes. Additionally, we have demonstrated a comparable ability to efficiently sort 25 modes in the mutually unbiased basis of angular position. Our technique will enable the development of high-dimensional QKD systems with record channel capacities of more than 4 bits per detected photon. In addition, this method will allow for detailed studies of fundamental quantum phenomena in the unbounded state-space of orbital angular momentum.

 \vspace{4mm} \noindent\normalsize\textbf{Methods}
\vspace{4mm}

\textbf{Equipments and experimental techniques.} A Helium Neon laser beam (632.8 nm) is spatially filtered using a single-mode fiber. The output of the fiber is collimated and illuminates a high-definition Holoeye PLUTO phase-only SLM. The OAM and ANG modes are created using computer generated holograms \cite{Arrizon} (See Supplementary Note 4). An aperture in used in between a 4f system to pick up the first order diffracted beam. Machined PMMA (Poly methyl methacrylate) refractive elements mounted in a Thorlabs cage system are used for log-polar coordinate mapping \cite{Lavery}. Two Holoeye PLUTO phase-only SLMs are used for realizing the hologram for the fan-out element and its corresponding phase-correcting element. A telescope system is used for coherent imaging of ANG modes to the fan-out hologram. Single images were obtained on a Canon 5D mark III  CCD camera providing 14 bits of dynamic range. A Semrock 632.8 nm MaxLine filter is used in front of the camera to block stray light.  An exposure time of 125 \(\mu\)s is chosen to further the suppress background light.

\vspace{4mm} \noindent\normalsize\textbf{Acknowledgements}
\vspace{4mm}

The authors would like to thank M. O'Sullivan, B. Rodenburg, M. Lavery, Dr.\,E. Karimi, Dr.\,M. Padgett, and Dr.\,D. Gauthier for helpful discussions. This work was supported by the DARPA InPho Program.  In addition, M. Malik acknowledges funding from the European Commission through a Marie Curie Fellowship, and RWB acknowledges support from the Canada Excellence Research Chairs program.

\vspace{4mm} \noindent\normalsize\textbf{Author Contributions}
\vspace{4mm}

M.Mirhosseini designed the experiment. M. Mirhosseini, M. Malik and Z. S. performed the experiment and analyzed data. R. W. B. supervised the project. M.Mirhosseini wrote the manuscript with contributions from all authors.

\vspace{4mm} \noindent\normalsize\textbf{Additional information}
\vspace{4mm}

The authors declare no competing financial interests. Reprints and permission information are available online at http://npg.nature.com/reprintsandpermissions/. Correspondence and requests for materials should be addressed to M. Mirhosseini.

\end{document}


\let\oldthebibliography=\thebibliography
\let\oldendthebibliography=\endthebibliography
\renewenvironment{thebibliography}[1]{%
    \oldthebibliography{#1}%
    \setcounter{enumiv}{28}%
}{\oldendthebibliography}


\vspace{4mm} \noindent\Large\textbf{Supplementary Figure S1}
\vfill
\begin{figure}[h!]
\centerline{\includegraphics[scale= 0.80]{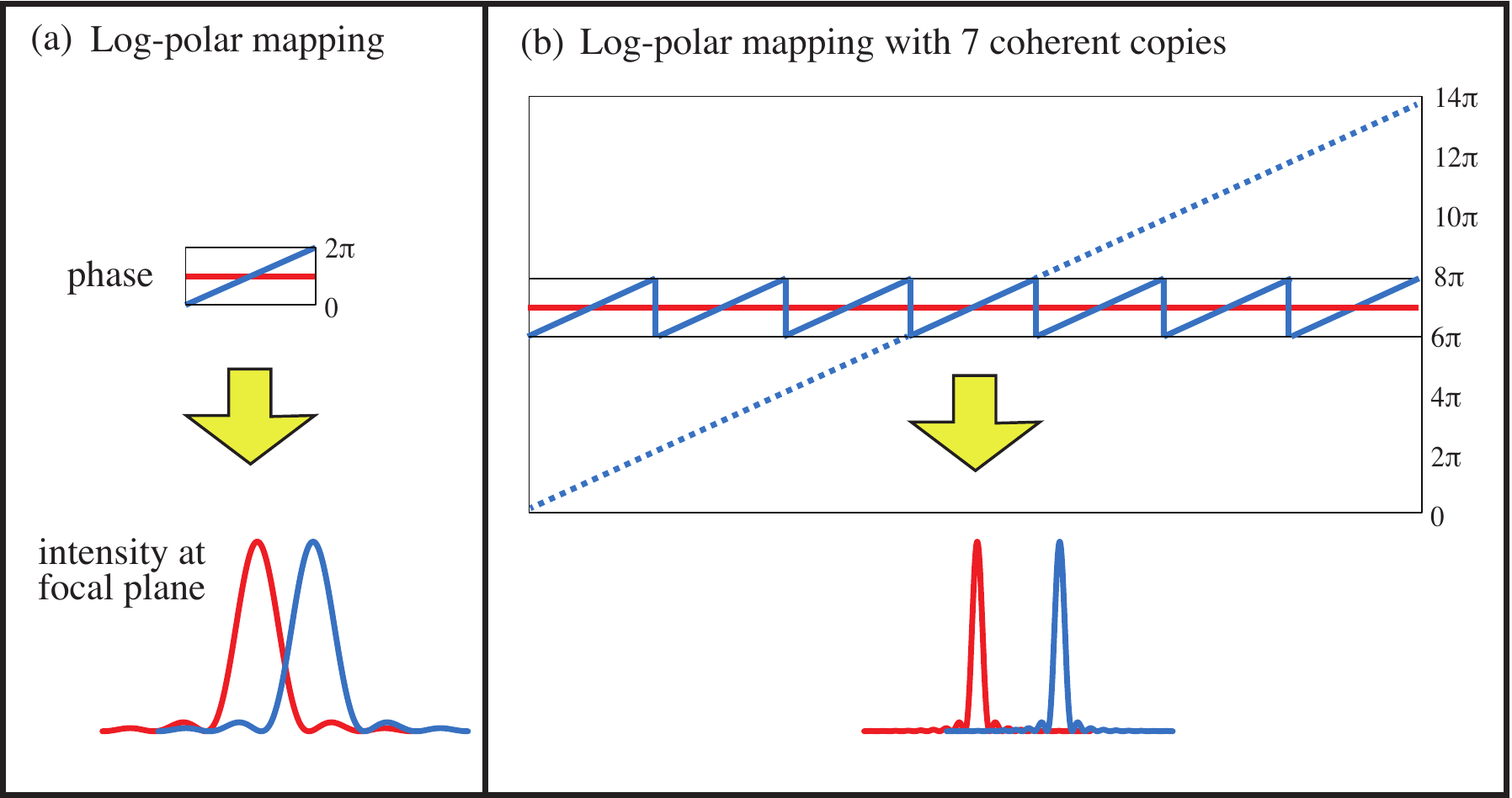}}
{\textbf{Supplementary Figure S1:} \textbf{Mode overlap between transformed OAM modes.} The phase structure of two neighboring OAM modes and the resulting intensity patterns in the Fourier domain for \textbf{a)} perfect log-polar mapping, and \textbf{b)} perfect log-polar mapping combined with beam copying.}
\label{fig:SepEf}
\end{figure}
\vfill
\clearpage

\vspace{4mm} \noindent\Large\textbf{Supplementary Figure S2}
\vfill
\begin{figure}[h!]
\centerline{\includegraphics[scale= 0.5]{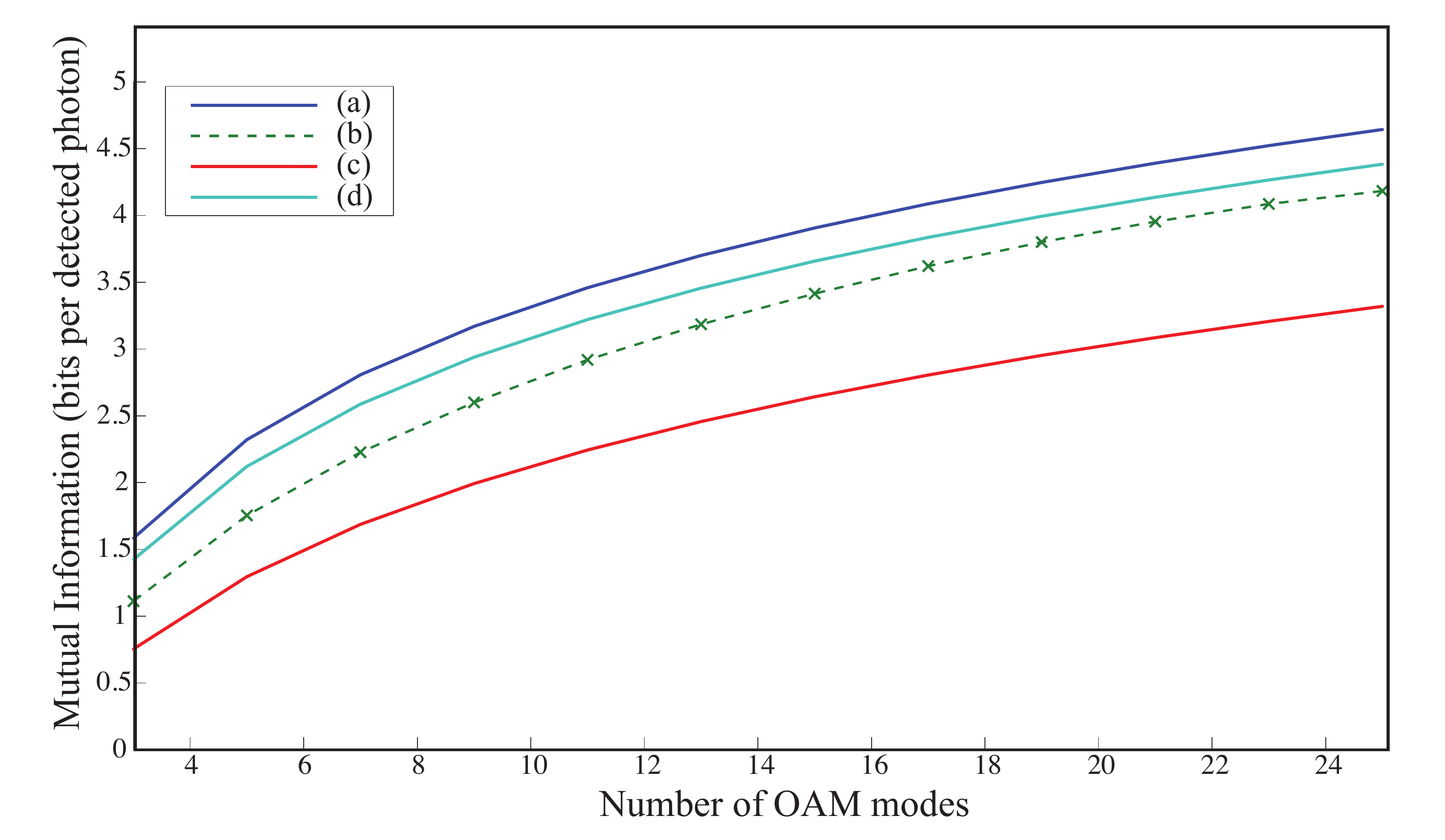}}
{\textbf{Supplementary Figure S2:} \textbf{Mutual information as a function of the number of the modes.} Mutual information as calculated for \textbf{a)} ideally separated modes, \textbf{b)} data from this experiment, \textbf{c)} perfect log-polar mapping, and \textbf{d)} perfect log-polar mapping combined with beam copying.}
\label{fig:Results-MI}
\end{figure}
\vfill
\clearpage

\vspace{4mm} \noindent\Large\textbf{Supplementary Figure S3}
\vfill
\begin{figure}[h!]
\centerline{\includegraphics[scale= 0.50]{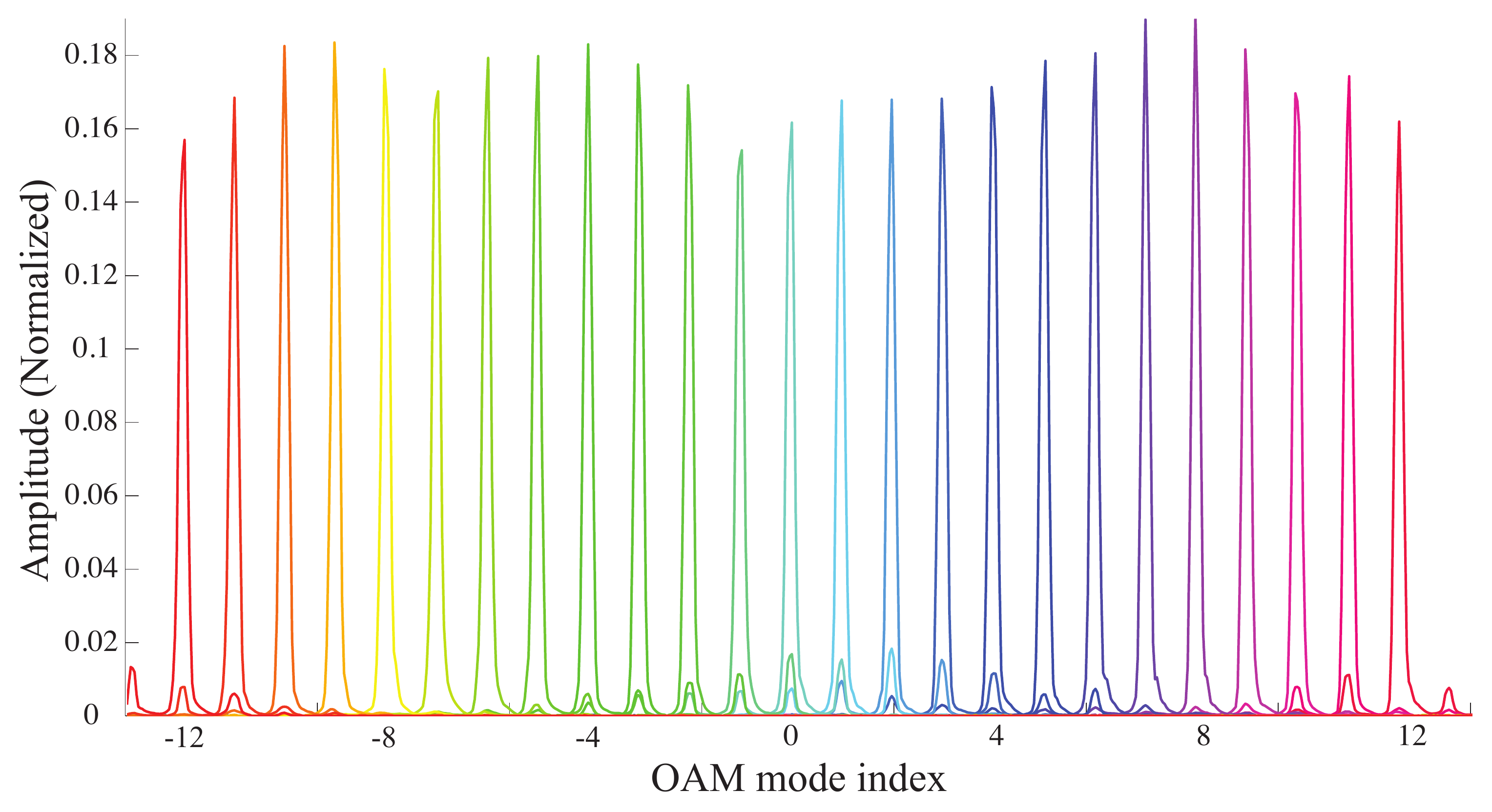}}
{\textbf{Supplementary Figure S3:} \textbf{Intensity profile of the transformed OAM modes.} The incident power in each OAM mode (summed along the vertical axis) as a function of the horizontal coordinate on the CCD camera. The center of each mode is labelled with the corresponding OAM mode index. The sum of the total power in each mode is normalized to unity.}
\label{fig:Results-OAM}
\end{figure}
 \vfill
 \clearpage

\vspace{4mm} \noindent\Large\textbf{Supplementary Figure S4}
\vfill 
\begin{figure}[h!]
\centerline{\includegraphics[scale= 0.50]{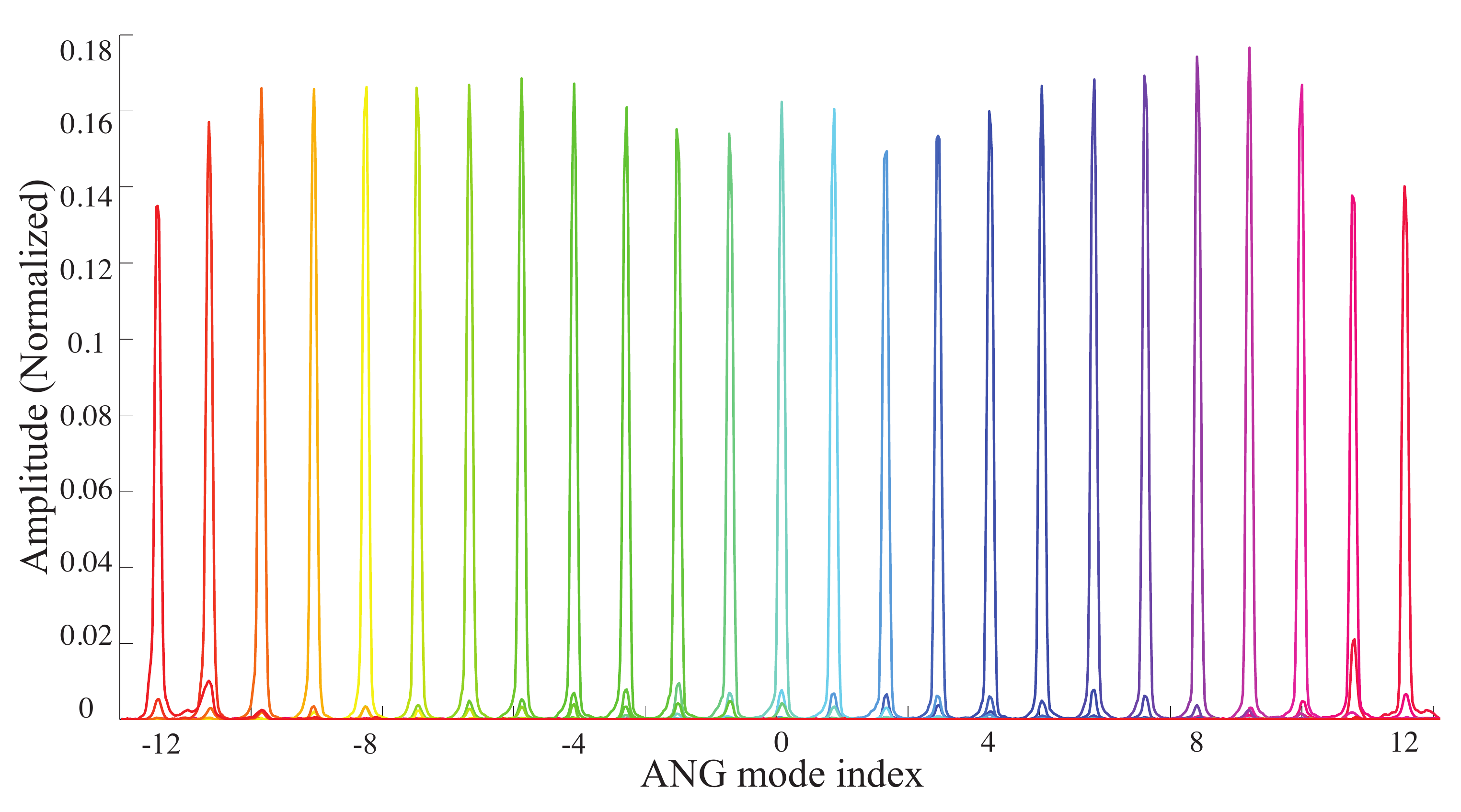}}
{\textbf{Supplementary Figure S4:} \textbf{Intensity profile of the transformed ANG modes.} The total incident power in each ANG mode as a function of the horizontal coordinate on the CCD camera. The center of each mode is labelled with the corresponding ANG mode index. Sum of the total power in each mode is normalized to unity.}
\label{fig:Results-ANG}
\end{figure}
 \vfill 
\clearpage

\vspace{4mm} \noindent\Large\textbf{Supplementary Table S1}
\vfill 
\begin{table*}[h!]
{\footnotesize\begin{tabular*}{1.0\textwidth}{@{\extracolsep{\fill}} l c c c c c c c}
\hline\hline
  Diffraction order \(m\) & -3 & -2 & -1 & 0 & 1 & 2 & 3\\
  \hline
  \\
\( \gamma_7(m)\) & 1.24 & 1.45 & 1.28 & 1 & 1.28 & 1.45 & 1.24\\
 \\
\(\alpha_7(m)\) (rad) & 7.03 & 1.89 & -0.99 & 0 & -0.99 & 1.89 & 7.03\\
\end{tabular*}}
%
{\footnotesize\begin{tabular*}{1.0\textwidth}{@{\extracolsep{\fill}} l c c c c c c c c c}
\hline
  Diffraction order \(m\) & -4 & -3 & -2 & -1 & 0 & 1 & 2 & 3 & 4\\
  \hline
  \\
\( \gamma_9(m)\) & 1.03 & 0.943 & 0.963 & 0.971 & 1 & 0.971 & 0.963 & 0.943 & 1.03\\
 \\
\(\alpha_9(m)\) (rad) & 1.41& 3.03 & 5.57 & 0.72 & 0 & 0.72 & 5.57 & 3.03 & 1.41 \\
\hline\hline
\end{tabular*}}

\vspace{1mm}
{\textbf{Supplementary Table S1:} \textbf{Parameters for fan-out elements.}  \textbf{(top)} Grating for creating seven copies. \textbf{(bottom)} Grating for creating nine copies. }
\end{table*}
 \vfill 
\clearpage

\vspace{4mm} \noindent\Large\textbf{Supplementary Note 1. Optimal Designs for Refractive Beam Copying }
\vspace{4mm}
\normalsize

Phase-only elements for beam copying can be designed using phase-retrieval and other iterative algorithms [25]. Alternatively, an analytical optimization process can be used to design such elements [26]. The results from these two methods agree closely. These elements have the general form
\begin{equation}
\Psi_{2N+1}(x) = \tan^{-1} \left(\frac{\sum\limits_{m=-N}^{N}\gamma_m \sin[(2\pi s/\lambda)mx+\alpha_m]}{\sum\limits_{m=-N}^{N}\gamma_m \cos[(2\pi s/\lambda)mx+\alpha_m]}\right).
\tag{S1}
\end{equation}
The values of \(\gamma\) and \(\alpha\) for fan-out elements creating 7 and 9 copies are listed in Supplementary Table S1.
\clearpage

\vspace{4mm} \noindent\Large\textbf{Supplementary Note 2. Separation Efficiency }
\vspace{4mm}
\normalsize

We define the separation efficiency as the probability of detecting an OAM mode correctly. The mode sorter maps different OAM modes to a series of spots which their centroids separated. 
Assuming a perfect log-polar mapping, each OAM mode is mapped to a plane-wave which has a finite size [12]. 
\begin{equation}
E(v ) = e^{i\ell v/a }\text{rect}(\frac{v}{2\pi a}).
\tag{S2}
\end{equation}
The separability of the form of these modes allows for a one-dimensional analysis. This unwrapped mode is focused to a spot. Using basic Fourier analysis, the spot pattern is found to be
\begin{equation}
E(x ) \propto \frac{\sin(\pi[ \frac{ ax}{\lambda f}-\ell])}{\frac{\pi x}{\lambda f}},
\tag{S3}
\end{equation}
where \(f\) is the focal length of the lens. Two transformed neighboring OAM modes are orthogonal in terms of their amplitude due the unitary nature of the involved transformation. The amplitude profile of these modes , however, is the familiar sinc form, which has a strong central components with side lobes around it (See Supplementary Fig. S1). As a result, there is an overlap between two neighboring OAM modes.

The probability of detecting each mode is equal to the portion of the intensity which falls into the corresponding spatial bin for an OAM mode. To treat all the OAM modes equally, the spatial bins have to be confined by the borders between the neighboring modes. Using the above analysis, the separation efficiency can be calculated as

\begin{equation}
\text{S. E.} = \frac{\int_{-1/2}^{1/2}{{\left|\frac{\sin(\pi x)}{\pi x}\right|}^2} dx}{\int_{-\infty}^{\infty}{{\left|\frac{\sin(\pi x)}{\pi x}\right|}^2} dx} = 0.774 = 77.4 \% . 
\tag{S4}
\end{equation}

 When a fan-out element is added, the unwrapped OAM mode is copied to \(N\) different copies. For an odd number of copies
 \begin{equation}
E(v ) = \frac{1}{\sqrt{N}} e^{i\ell v/a } \sum_{i = -(N-1)/2}^{(N-1)/2} \text{rect}(\frac{v - ia}{a}) = \frac{1}{\sqrt{N}} e^{i\ell v/a } \text{rect}(\frac{v}{Na}).
\tag{S5}
\end{equation}
The factor of \(1/\sqrt{N}\) has to be added for conservation of energy. This extended unwrapped mode is focused to a spot which is \(N\) times narrower
\begin{equation}
E(x ) \propto  \frac{\sqrt{N}\sin(\pi[ \frac{Nxa}{\lambda f} -\ell])}{\frac{\pi x}{\lambda f}}.
\tag{S6}
\end{equation}

In this case, the sinc functions are \(N\) times narrower as compared to the previous case but the spacing between them has not changed (See Supplementary Fig. S1).
We can use an analysis similar to above to calculate the separation efficiency for the case where the log-polar mapping is enhanced with beam copying. For the case of seven copies, the separation efficiency is equal to 

\begin{equation}
\text{S. E.} = \frac{\int_{-1/2}^{1/2}{{\left|\frac{\sin(7\pi x)}{7\pi x}\right|}^2} dx}{\int_{-\infty}^{\infty}{{\left|\frac{\sin(7\pi x)}{7 \pi x}\right|}^2} dx} = 0.974 = 97.4 \% . 
\tag{S7}
\end{equation}
\clearpage

\vspace{4mm} \noindent\Large\textbf{Supplementary Note 3. Effects of Cross-talk and Loss on Mutual Information }
\vspace{4mm}
\normalsize

A common measure to quantify the degree of separation of modes in a communication system is the mutual information between the sent and received modes. This quantity is defined as
\begin{equation}
I (Y,X) = \sum_{i,j} P(y_j,x_i) \log_2\left[\frac{P(y_j,x_i)}{P(x_i)P(y_j)} \right].
\tag{S8}
\label{eq:MI}
\end{equation}

Here, \(x_i\) is the event of sending a photon with the OAM value of \(\ell_i\) and \(y_j\) is the event of detecting a photon with the OAM value of \(\ell_j\). Assuming a uniform probability for sending \(N\) modes, \(P_x\) equals \(1/N\) for all \(x\). The joint probability of detection \(P(y_j,x_i)\) can be calculated from the conditional probability \(P(y_j | x_i)\) by using the Bayes' rule 

\begin{equation}
P(y_j,x_i) = P(y_j | x_i) P(x_i)  = \frac{1}{N} P(y_j | x_i) .
\tag{S9}
\end{equation}

We have spatially binned the detector plane in order to detect each mode. In this case, the conditional probability \(P(y_j | x_i)\) is proportional to incident power from mode \(i\) to the bin corresponding to the mode \(j\) (For patterns of the detected modes, see Supplementary Fig. S3 and Supplementary Fig. S4). The proportionality constant can be simply found by imposing the probability normalization condition \( \sum_{j} P(y_j | x_i) = P(x_i) = \frac{1}{N}\). 

The normalization condition described above excludes the loss in the channel (we denote the conditional probabilities measured from this method as  \(P(y_j | x_i) = S_{ji}\)). As a result, the mutual information calculated from this method only quantifies the degree of cross-talk and it is usually measured in units of bits per detected photon. Since our method relies on phase-only optical elements, in principle it can achieve a power transmission of unity. It should be noted that although the fan-out elements have a smaller-than-unity efficiency (99.3 for 9 copies and 96.7 for 7 copies) when considering only the desired copies, they are phase-only components and are strictly lossless. In our experiment and the theory work published previously [24] we collect all the light diffracted from the fan-out element and as a result there is no inherent loss in our method.

Using the values of \(S_{ji}\) matrix, the mutual information between the transmitted and the detected modes can be calculated as
\begin{equation}
I  (\text{bits per detected photon}) = \frac{1}{N} \sum_{{i,j} =1}^N S_{ji} \log_2\left[\frac{N S_{ji}}{\sum_{i} S_{ji}} \right].
\tag{S10}
\end{equation}

We have used the conditional probability matrix \(S_{ji}\) to calculate the mutual information for OAM basis sets with different dimensions (Supplementary Fig. S2). For comparison,  we have also plotted the theoretical upper bound for the mutual information of a system employing \(N\) OAM modes, which equals \(\log_2(N)\) (Supplementary Fig. S2). Supplementary Fig. S2 also includes values of mutual information calculated from theory for a perfect log-polar coordinate mapping [12] (c) and perfect log-polar mapping combined by beam copying (d) [24] (all values are in bits per detected photons).

To achieve a complete characterization of the information capacity of a channel, one must take into account the loss as well as the cross-talk in the calculation of the mutual information between the transmitted and the detected modes. Due to the presence of loss in a realistic communication channel, sometimes the transmitted photons lead to no detection event in the receiver's side. In this situation, the failure to detect any mode should be treated as an alternative possible event for the receiver \cite{Power}. However, it is easy to check that if the value of the loss is equal for all the modes, the no-detection event contains no information about the transmitted mode and hence it can be discarded from the calculation of the mutual information. In this case, the only effect of loss is in reducing the probability of successfully detecting each mode.

If the transmitted photon is lost in the channel by a probability of \(P_{L}\), a successful transmission will have a probability of \(P_{T}\), where \(P_{T} = 1- P_{L}\). In this case, the joint probability of detection \(P(y_j,x_i)\) for each mode is calculated as 

\begin{equation}
P(y_j,x_i) = P(y_j | x_i) P(x_i)  = \frac{1}{N} P_{T} S_{ji}, 
\tag{S11}
\end{equation}
where \(S_{ji}\) are measured as before. Substituting the joint probability values in Eq. (S8) we have 

\begin{equation}
I  (\text{bits per launched photon}) = \frac{1}{N} \sum_{{i,j} =1}^N P_{T} S_{ji} \log_2\left[\frac{N \cancel{P_{T}} S_{ji}}{\sum_{i} \cancel{P_{T}}  S_{ji}} \right] .
\tag{S12}
\end{equation}
This can be further simplified as
\begin{equation}
I  (\text{bits per launched photon}) = P_{T} \times I  (\text{bits per detected photon}).
\tag{S13}
\end{equation}
This relation has a simple interpretation.\,It states that the mutual information between the transmitted and the detected modes is scaled down by the the overall power transmission efficiency of the optical communication link. 

We have measured the overall power efficiency of our system to be about 15 percent. Using this number, our system transfers 4.18 bits per detected photon and \(4.18 \times 0.15 = 0.63\) bits per launched photon for OAM modes. Similarly, the ANG modes are transferred and separated with a mutual information of 4.16 bits per detected photon and \(4.16 \times 0.15 = 0.62\) bits per launched photon.  However, it should be emphasized once more that the loss in our system is a consequence of the specific experimental realization we have utilized, and is not inherent to our sorting method. Straightforward technological improvements could lead to a significant increase in the number of bits per launched photon. Manufacturing of the fan-out and its phase corrector in the form of refractive elements can achieve power transmission efficiencies close to a hundred percent. In fact, fan-out elements with efficiencies of more than 95\% are commercially available. Alternatively, the required elements can be realized using new generation of SLMs which can achieve diffraction efficiencies of more than 90\%.
\clearpage

\vspace{4mm} \noindent\Large{\textbf{Supplementary Note 4. Mutually Unbiased Bases and Angular Modes}}
\vspace{4mm}

\normalsize
To guarantee security in the BB84 protocol, at least two mutually unbiased bases (MUBs) of polarization are needed [4]. Similarly, a basis unbiased with respect to the basis of orbital angular momentum (OAM) is needed in an OAM-based quantum key distribution (QKD) system [10]. Such a basis can be constructed using the following formula

\begin{equation}
\label{eq:angMode}
|\theta_j\rangle= \frac{1}{\sqrt{2L+1}} \sum_{\ell=-L}^{L} |\ell\rangle \: e^{-i 2 \pi j\ell/(2L+1)}.
\tag{S14}
\end{equation}
These modes are referred to as the angular (ANG) modes due to the angular confinement of their intensity pattern. Some of the ANG modes made from superposition of \(\ell = -12\) to \(\ell = -12\) can be seen in Fig. 4.

We have used a phase-only spatial light modulator (SLM)  to realize holograms for generating OAM and ANG modes. In order to generate a spatial mode described by \(a(x,y)e^{i\phi(x,y)}\), the following phase pattern is impressed on the SLM
\begin{equation}
\Psi(x,y) = \left(1-\text{sinc}^{-1}[a(x,y)]\right)\phi(x,y).
\tag{S15}
\end{equation}
A blaze function is usually added to the phase of the target beam to achieve better separation of the desired and undesired parts in the Fourier domain [27]. Fig. 4 shows some of the holograms used for generating OAM and ANG modes.
\clearpage

\renewcommand{\refname}{}
\vspace{4mm} \noindent\Large\textbf{Supplementary References }